\documentclass[aps,prl,twocolumn,superscriptaddress,showpacs,preprintnumbers,amsmath,amssymb]{revtex4}
\usepackage{graphicx} 
\usepackage{dcolumn}  
\usepackage{bm}
\usepackage{amsmath}
\usepackage{amssymb}

\graphicspath{{ps}}

\begin{document}



\title{ \quad\\[1.0cm] Search for $B_{s}^{0}$ $\to$ $hh$ decays at the 
$\Upsilon(5S)$ resonance} 

\affiliation{Budker Institute of Nuclear Physics, Novosibirsk}
\affiliation{Faculty of Mathematics and Physics, Charles University, Prague}
\affiliation{University of Cincinnati, Cincinnati, Ohio 45221}
\affiliation{Department of Physics, Fu Jen Catholic University, Taipei}
\affiliation{The Graduate University for Advanced Studies, Hayama}
\affiliation{Hanyang University, Seoul}
\affiliation{University of Hawaii, Honolulu, Hawaii 96822}
\affiliation{High Energy Accelerator Research Organization (KEK), Tsukuba}
\affiliation{Institute of High Energy Physics, Chinese Academy of Sciences, Beijing}
\affiliation{Institute of High Energy Physics, Vienna}
\affiliation{Institute of High Energy Physics, Protvino}
\affiliation{Institute for Theoretical and Experimental Physics, Moscow}
\affiliation{J. Stefan Institute, Ljubljana}
\affiliation{Kanagawa University, Yokohama}
\affiliation{Institut f\"ur Experimentelle Kernphysik, Karlsruher Institut f\"ur Technologie, Karlsruhe}
\affiliation{Korea Institute of Science and Technology Information, Daejeon}
\affiliation{Korea University, Seoul}
\affiliation{Kyungpook National University, Taegu}
\affiliation{\'Ecole Polytechnique F\'ed\'erale de Lausanne (EPFL), Lausanne}
\affiliation{Faculty of Mathematics and Physics, University of Ljubljana, Ljubljana}
\affiliation{University of Maribor, Maribor}
\affiliation{Max-Planck-Institut f\"ur Physik, M\"unchen}
\affiliation{University of Melbourne, School of Physics, Victoria 3010}
\affiliation{Nagoya University, Nagoya}
\affiliation{Nara Women's University, Nara}
\affiliation{National Central University, Chung-li}
\affiliation{National United University, Miao Li}
\affiliation{Department of Physics, National Taiwan University, Taipei}
\affiliation{H. Niewodniczanski Institute of Nuclear Physics, Krakow}
\affiliation{Nippon Dental University, Niigata}
\affiliation{Niigata University, Niigata}
\affiliation{Novosibirsk State University, Novosibirsk}
\affiliation{Osaka City University, Osaka}
\affiliation{Panjab University, Chandigarh}
\affiliation{University of Science and Technology of China, Hefei}
\affiliation{Seoul National University, Seoul}
\affiliation{Sungkyunkwan University, Suwon}
\affiliation{School of Physics, University of Sydney, NSW 2006}
\affiliation{Tata Institute of Fundamental Research, Mumbai}
\affiliation{Excellence Cluster Universe, Technische Universit\"at M\"unchen, Garching}
\affiliation{Toho University, Funabashi}
\affiliation{Tohoku Gakuin University, Tagajo}
\affiliation{Tohoku University, Sendai}
\affiliation{Department of Physics, University of Tokyo, Tokyo}
\affiliation{Tokyo Metropolitan University, Tokyo}
\affiliation{Tokyo University of Agriculture and Technology, Tokyo}
\affiliation{IPNAS, Virginia Polytechnic Institute and State University, Blacksburg, Virginia 24061}
\affiliation{Yonsei University, Seoul}
  \author{C.-C.~Peng}\affiliation{Department of Physics, National Taiwan University, Taipei} 
 \author{P.~Chang}\affiliation{Department of Physics, National Taiwan University, Taipei} 
  \author{I.~Adachi}\affiliation{High Energy Accelerator Research Organization (KEK), Tsukuba} 
  \author{H.~Aihara}\affiliation{Department of Physics, University of Tokyo, Tokyo} 
  \author{T.~Aushev}\affiliation{\'Ecole Polytechnique F\'ed\'erale de Lausanne (EPFL), Lausanne}\affiliation{Institute for Theoretical and Experimental Physics, Moscow} 
  \author{T.~Aziz}\affiliation{Tata Institute of Fundamental Research, Mumbai} 
  \author{A.~M.~Bakich}\affiliation{School of Physics, University of Sydney, NSW 2006} 
  \author{V.~Balagura}\affiliation{Institute for Theoretical and Experimental Physics, Moscow} 
  \author{E.~Barberio}\affiliation{University of Melbourne, School of Physics, Victoria 3010} 
  \author{K.~Belous}\affiliation{Institute of High Energy Physics, Protvino} 
% ...
\collaboration{The Belle Collaboration}


\begin{abstract}

We have searched for $B_{s}^{0} \to hh$ decays, where $h$ stands for a charged or neutral kaon, or a charged pion.
These results are based on a 23.6 fb$^{-1}$ data sample collected with the Belle detector on the $\Upsilon(5S)$ resonance at the KEKB asymmetric-energy $e^{+}$$e^{-}$ collider, containing 1.25 million $B_{s}^{(*)}\bar{B}_{s}^{(*)}$ events.  
We observe the decay $B_{s}^{0} \to K^{+}K^{-}$ and measure its branching fraction,
$\mathcal{B}(B_{s}^{0} \to K^{+}K^{-}) = [3.8_{-0.9}^{+1.0}(\mathrm{stat})\pm 0.5(\mathrm{syst})\pm 0.5(f_s)] \times 10^{-5}$. The first error is statistical,
the second is systematic, and the third error is due to the uncertainty in the 
$B^0_s$ production fraction in  $e^+e^-\to b\bar{b}$ events.   
No significant signals are seen in other decay modes, and we set  
upper limits at the 90\% confidence level:
$\mathcal{B}(B_{s}^{0} \to K^-\pi^{+})< 2.6 \times 10^{-5}$,
$\mathcal{B}(B_{s}^{0} \to \pi^{+}\pi^{-})< 1.2 \times 10^{-5}$ and
$\mathcal{B}(B_{s}^{0} \to K^0\bar{K}^0) < 6.6\times 10^{-5}$.

\end{abstract}

\pacs{
13.25.Hw, 14.40.Nd
}
\maketitle


{\renewcommand{\thefootnote}{\fnsymbol{footnote}}}
\setcounter{footnote}{0}




The recent observation of a significant difference between direct CP violation 
in $B^0\to K^\pm\pi^\mp$ and $B^\pm\to K^\pm \pi^0$~\cite{acp,qcon} was  
unexpected and has generated much discussion.
Possible explanations for this difference include a 
large color-suppressed tree amplitude~\cite{ctree}, new physics in the 
electroweak penguin loop~\cite{new}, or both~\cite{both}. 
Similar measurements of charmless two-body $B_s^0$ decays may
provide additional insight into this and other aspects of $B$ decays.
For instance, a comparison of the $CP$ violating asymmetries between the $B^0$ and $B_s^0$ may discriminate among new physics models~\cite{newbs};
the angles $\phi_1(\beta)$ and $\phi_3(\gamma)$ of the unitarity triangle may be extracted using the time evolution of the decays $B^0\to \pi^+\pi^-$ and $B_s^0\to K^+K^-$~\cite{plb459}; 
the branching 
fractions and $CP$ violating asymmetries of these two decays provide information on  $U$-spin symmetry breaking~\cite{PRL97061801}; and
the decay $B_s^0\to K^-\pi^+$ can be 
used to determine $\phi_3(\gamma)$~\cite{plb482}. 

The decay $B_s^0 \to K^+K^-$ is  of particular interest 
because its branching 
fraction is expected to be large, in analogy to that of $B^0\to K^+\pi^-$, and 
the 
final state is a $CP$ eigenstate. 
The time-dependent $CP$ asymmetry of this decay is sensitive to
the $B^0_s-\bar{B}_s^0$ mixing phase 
($\phi_s$) and the width difference of the two $B^0_s$ mass eigenstates ($\Delta\Gamma_s$);
these two parameters provide a clean probe of new physics beyond the 
Standard Model. CDF and D$\O$ have performed a time-dependent $CP$ analysis 
using $B^0_s\to J/\psi \phi$ events to  
 measure $\phi_s$ and $\Delta \Gamma_s$. The results are limited by statistics 
 and no significant deviations from the SM expectation are observed~\cite{phis}. 


Experimental results to date on charmless  $B_s^0$ decay have been 
limited to just a few measurements from CDF~\cite{CDFresult1,cdf,cdf2} and Belle~\cite{Belleresult}. 
In this paper, we report on a search for $B^0_s$ decays to $K^+K^-$, $K^0\bar{K}^0$, $K^-\pi^+$ and $\pi^+\pi^-$ based on a (23.6$\pm$0.3) 
fb$^{-1}$ $(L_{\rm int})$ data sample collected at the $\Upsilon$(5S) resonance with
the Belle detector operated at the KEKB asymmetric-energy (3.6 GeV on 8.2 GeV) $e^{+}e^{-}$ collider~\cite{KEKB}. 
In an earlier study, half of the center-of-mass (c.m.)  energy 
 was measured to be $E_{\rm beam}^{*} = (5433.5 \pm 0.5)$ MeV~\cite{5S1S}. 
At this energy, the total cross section for production of light quark pairs of 
the first 
two families is around 2.446 nb \cite{eeqq} while  
the cross section for $b\bar b$ events is 
$\sigma^{\Upsilon(5S)}_{b\bar{b}} = (0.302 \pm 0.014)$~nb, of which a fraction $f_s$ = $(19.5^{+3.0}_{-2.3})\%$ contains $B_s^0$ mesons \cite{crosssection}. 
Three production modes are kinematically allowed: 
$B_{s}^{0}\bar{B}_{s}^{0}$, $B_{s}^{*}\bar{B}_{s}^{0}$ and $B_{s}^{*}\bar{B}_{s}^{*}$, 
where the fraction of $B_{s}^{*}\bar{B}_{s}^{*}$ is $f_{B_s^*\bar{B}_s^*}$ = $(90.1^{+3.8}_{-4.0}\pm 0.2)\%$ ~\cite{dspi}.
The number of $B_s^*\bar{B}_s^*$ pairs is thus
computed as $N_{B_s^*\bar{B}_s^*}$ = $L_{\rm int} \times \sigma^{\Upsilon(5S)}_{b\bar{b}} \times f_s \times f_{B_s^*\bar{B}_s^*} = (1.25 \pm 0.19) \times 10^6$.

The Belle detector is a large-solid-angle magnetic
spectrometer that consists of a silicon vertex detector (SVD),
a 50-layer central drift chamber (CDC), an array of
aerogel threshold Cherenkov counters (ACC),  
a barrel-like arrangement of time-of-flight
scintillation counters (TOF), and an electromagnetic calorimeter
comprised of CsI(Tl) crystals (ECL) located inside 
a superconducting solenoid coil that provides a 1.5~T
magnetic field.  An iron flux-return located outside of
the coil is instrumented to detect $K_L^0$ mesons and to identify
muons (KLM).  The detector
is described in detail elsewhere~\cite{Belle}.

Charged kaons and pions are required to have a distance of closest approach
to the interaction point (IP) of less than 3.0 cm in the beam direction and   
less than 0.3 cm in the transverse plane. 
Charged kaons and pions are identified
using $dE/dx$ measurements from the CDC, Cherenkov light yields in the ACC,
and timing information from the TOF. This information is combined in a 
likelihood ratio,  $\mathcal{R}_{K/\pi}$ = $\mathcal{L}_{K}$/($\mathcal{L}_{\pi}$ + $\mathcal{L}_{K}$), where $\mathcal{L}_{K}$ $(\mathcal{L}_{\pi})$ 
is the likelihood that the track is a kaon (pion). 
Charged tracks with $\mathcal{R}_{K/\pi}>0.6$ are treated as kaons, and with 
$\mathcal{R}_{K/\pi}<0.6$ as
pions\cite{kid}. Furthermore, charged tracks positively identified as electrons or muons~\cite{kid} are 
rejected. 
With these selections, the kaon (pion) identification efficiency is about 83\% (88\%), while 12\% (8\%) of
kaons (pions) are misidentified as pions (kaons).
Neutral kaons are reconstructed in the $K_{S}^{0}$ $\to$ $\pi^{+}\pi^{-}$ decay 
channel and are required to have an invariant mass 
in the range 490 MeV/$c^{2}$ $<$ $M_{\pi^{+}\pi^{-}}$ $<$ 510 MeV/$c^{2}$. The intersection point of the $\pi^{+}\pi^{-}$ 
pair must be displaced from the IP~\cite{kfchenks}.

$B^0_{s}$ candidates are selected by combining kaons and pions in appropriate pairs 
and separated from background using two variables: the beam-energy-constrained 
mass, 
$M_{\mathrm{bc}} = \sqrt{E_{\mathrm{beam}}^{*2}-p_{B}^{*2}}$, and the energy difference,
$\Delta E = E_{B}^{*} - E_{\mathrm{beam}}^{*}$, where 
 $p_{B}^{*}$ and $E_{B}^{*}$ are the momentum and energy of the reconstructed 
$B^0_{s}$ meson in the c.m. frame, respectively.  Figure 1 shows the 
GEANT-based~\cite{geant} Monte Carlo $\Delta E$-$M_{\rm bc}$ distributions for the $B^0_{(s)} \to hh$ candidates from various two-body, three-body and 
four-body $\Upsilon(5S)$ decays generated with a $B$ meson decaying 
into an $hh$ pair. Although only one $B$ meson per event is fully 
reconstructed, we can identify the $\Upsilon(5S)$ decay from which it 
originates
based on its location in the $\Delta E$-$M_{\rm bc}$ plane.  
Candidates with 
$-0.2$  GeV $<\Delta E<0.2$ GeV
and $5.35$ GeV/$c^{2}$ $< M_{\mathrm{bc}} < 5.45$ GeV/$c^{2}$ are selected.
Since the dominant source of $B_s^0$ mesons is  
$\Upsilon(5S)\to B_{s}^{*}\bar{B}_{s}^{*}$,  
we search for  $B^0_{s}$ 
mesons only in this decay channel 
and define the signal region to be $-0.1$ GeV $<\Delta E < 0.0$ GeV and $5.40$ GeV/$c^{2}$ $< M_{\mathrm bc} < 5.43$ GeV/$c^{2}$.

After applying the $M_{\mathrm{bc}}$-$\Delta E$ selection, there are 14528, 
30613, 27454, and 444 candidates for the $K^+K^-, K^-\pi^+, \pi^+\pi^-$ and 
$K^0 \bar{K}^0$ modes, respectively. These candidates are predominantly from 
continuum events, {\it i.e.}, $e^{+}e^{-} \to q\bar{q}$, where $q$ stands for a $u$, $d$, $s$ or $c$ quark.
The event topology difference between $q\bar{q}$ and $b\bar{b}$ events is
exploited by computing 
 a Fisher discriminant~\cite{fisher} based on a set of modified Fox-Wolfram moments ~\cite{KSFW}.
Signal ($\mathcal{L}_{\rm{s}}$) and background ($\mathcal{L}_{q\bar{q}}$) likelihoods are formed 
using a 
Monte Carlo (MC) simulation and data outside the signal region, respectively. 
They are combined into 
a likelihood ratio  
$\mathcal{R} = \mathcal{L}_{\rm{s}}/(\mathcal{L}_{\mathrm{\rm s}}+\mathcal{L}_{q\bar{q}}$). 
The selection criterion, based on $\mathcal{R}$, is determined by maximizing 
$S/\sqrt{S+B}$, where
$S$ and $B$ are the number expected in the signal region of signal or background events, respectively. 
The expected signals are determined by assuming the following branching 
fractions \cite{hhbr}:
${\cal B}(B^0_s\to K^+K^-)=2.6\times 10^{-5}$,
${\cal B}(B^0_s\to K^-\pi^+)=4.6\times 10^{-6}$,
${\cal B}(B^0_s\to K^0\bar K^0)=1.2\times 10^{-5}$,
${\cal B}(B^0_s\to \pi^+\pi^-)=1.0\times 10^{-7}$.
For the $B_s^0 \to K^+K^-$ mode, we apply a looser criterion on $\cal{R}$ if the event contains an identified electron (muon) with
momentum larger than 0.5 (0.8) GeV/$c$. After the $\mathcal{R}$ requirement,
300, 444, 188 and 345 candidates are retained for the $K^+K^-, K^-\pi^+, \pi^+\pi^-$,
and $K^0\bar{K}^0$ modes, respectively.

\begin{figure}[t!]
\begin{center}
\hspace{-0.2cm}
\includegraphics[width= 8.4cm]{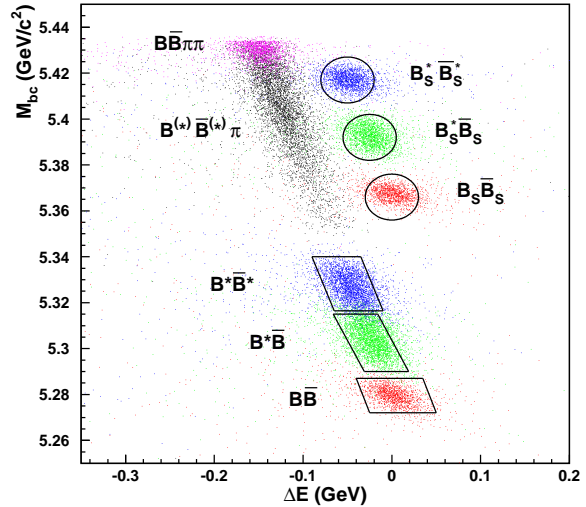}\\
\end{center}
\caption{
Monte Carlo distributions of $\Delta E$-$M_{\rm bc}$ for $B^0_{(s)}\to hh$
candidates from various $\Upsilon(5S)$ decay modes with $B$ mesons. 
Events in the circles are from
$\Upsilon(5S)\to B^{0(*)}_s \bar{B}^{0(*)}_s$; candidates in the 
parallelograms are generated with $\Upsilon(5S)\to B^{0(*)} \bar{B}^{0(*)}$; 
 three-body $B^{(*)}\bar{B}^{(*)}\pi$ and four-body $B\bar{B}\pi \pi$ events
are located at $M_{bc}> 5.35$ GeV/$c^2$ and $\Delta E<-0.05$ GeV.     
}
\label{fig:bbmc}
\end{figure}

Backgrounds from $B$ meson decays are studied using large MC samples, which
include  $\Upsilon(5S)\to B^{(*)}_s\bar{B}^{(*)}_s$, $\Upsilon(5S)\to B^{*}\bar{B}\pi$ and $\Upsilon(5S) \to B\bar{B}\pi\pi$ events.  The contributions from
$\Upsilon(5S)\to B\bar{B}, \Upsilon(5S)\to B^*\bar{B}$ and $ \Upsilon(5S)\to 
B^*\bar{B^*}$ are negligible since the $hh$ candidates from the corresponding 
$B$ decays lie outside the 
required $M_{\rm bc}$-$\Delta E$ region as shown in Fig.~\ref{fig:bbmc}. 
Out of the four $B^0_s$ decays, $B$ meson backgrounds only appear in the 
$B^0_s \to K^-\pi^+$ mode. A non-negligible contribution from $\Upsilon(5S)\to B^{(*)}_s\bar{B}^{(*)}_s$ events is present when one of the kaons from $B^0_s  \to K^+ K^-$ is misidentified as a pion (cross-feed). The second $B$ meson 
background is the $\bar{B}^0 \to K^-\pi^+$ events from three-body $\Upsilon(5S)\to B^{*}\bar{B}\pi$ and four-body
$\Upsilon(5S) \to B\bar{B}\pi\pi$ decays. With
 the branching fractions of  $\Upsilon(5S)\to B^{*}\bar{B}\pi$ and $\Upsilon(5S) \to B\bar{B}\pi\pi$ assumed to
be 6.8\% and 9.2\%, respectively~\cite{5SBBPi}, we expect to reconstruct about 
five  $\bar{B}^0 \to K^-\pi^+$ events, located outside the signal region. These
cross-feed and  $\bar{B}^0 \to K^-\pi^+$ backgrounds  are 
considered when extracting the $B^0_s\to K^-\pi^+$ signals.

We perform an unbinned extended maximum likelihood 
fit to 
$M_{\mathrm{bc}}$ and $\Delta E$ to extract signal yields.  The likelihood function is defined as :\\
\begin{equation}
\mathcal{L}  = {e^{-\sum_{j} N_{j}} \over N!}\prod_{i=1}^{N} 
\mathstrut^{\mathstrut}_{\mathstrut} \sum_{j} N_{j} P_{j}, 
\end{equation}
where  $N$ is the total number of events, 
$i$ runs over the selected events and $j$ over the signal and background components.
$N_{j}$ is the number of events for component $j$,
and $P_{j}$ is the corresponding probability density function (PDF). The 
continuum PDF is the product of a second-order polynomial function for $\Delta E$ and an
empirical ARGUS function~\cite{argus} for $M_{\rm bc}$.
For each mode, the signal PDF is modeled from MC with a Gaussian function for $M_{\mathrm{bc}}$ and a double Gaussian for $\Delta E$. 
The mean values of $M_{\mathrm{bc}}$ and $\Delta E$ are calibrated with $B^0_{s}\to D_{s}^{+}\pi^{-}$ decays, and the $\Delta E$ width is calibrated with $\bar{D}^0 \to K^{+} \pi^{-}$ decays. 
For the $B_s^0 \to K^-\pi^+$ mode, 
the $B_s^0 \to K^+K^-$ cross-feed and the $\bar{B}^0 \to K^-\pi^+$ background are modeled by 
two-dimensional smoothed histogram functions.
Yields for signal and continuum candidates, and the parameters of the continuum PDF, are allowed to float in the fit while the parameters for
other components are fixed. 
The branching fraction ($\mathcal{B}$) is computed as:
\begin{equation}
\mathcal{B}
= {N_s\over \epsilon \times 2N_{B_{s}^{*}\bar{B}_{s}^{*}}} ,
\end{equation}
where $N_s$ is  the fitted signal yield and $\epsilon$ is the MC efficiency. 



Two types of systematic uncertainties are considered:  uncertainties associated with the fit and uncertainties on the
signal reconstruction efficiency and number of $B_s^0$ meson pairs.
The fit systematic uncertainties are due to the modeling of the signal and continuum PDFs, and the statistical uncertainties in the background yields that were
 fixed in the fit.
The uncertainties due to the signal PDFs are obtained by varying each
PDF parameter successively by one standard deviation and repeating the fit.
The systematic uncertainty is the quadratic sum of the changes in the 
signal yield.
The uncertainty in modeling the continuum background 
is studied by changing the $\Delta E$ PDFs from  second- to 
 first-order polynomials. 
For the $B^0_s \to K^-\pi^+$ mode, the fit is repeated with the $B^0_s\to 
K^+K^-$ cross-feed yield varied by plus or minus one standard deviation and 
the signal yield variations are assigned as systematic uncertainties.  The 
systematic error that arises from the $\bar{B}^0 \to K^-\pi^+$ background is
obtained by taking  the difference of the signal yield with and without 
including the $\bar{B}^0 \to K^-\pi^+$ PDF in the fit.

\begin{table}[htb]
\caption{Contributions to the systematic error (\%). }

\label{sys1}
\begin{tabular}
{@{\hspace{0.1cm}}l@{\hspace{0.2cm}}  @{\hspace{0.2cm}}cccc@{\hspace{0.1cm}}}
\hline \hline
Source & $K^{+}K^{-}$ & $K^{-}\pi^{+}$ & $\pi^{+}\pi^{-}$ & $K^{0}K^{0}$ \\
\hline
Signal PDF & $ 2.3$ & $ 10.6$ & $ 10.3$ & $ 6.8$\\
Continuum PDF & $ 0.7$ & $ 1.5$ & $ 3.9$ & $ 6.3$ \\
Cross-feed background &  -- & $ 5.5$ & -- & -- \\
$\bar{B}^0 \to K^-\pi^+$ background & -- & $ 7.1$ & -- & -- \\
$\mathcal{R}$ requirement & $ 12.0$ & $ 12.8$ & $ 16.5$ & $ 4.8$ \\
$\mathcal{R}$$(K/\pi)$ requirement & $ 1.4$ & $ 1.4$ & $ 1.3$ &  -- \\
$K_{S}^{0}$ reconstruction & -- & -- & -- & $ 9.8$ \\
Track reconstruction  & $ 2.0$ & $ 2.0$ & $ 2.0$ & $ 0.0$ \\
$\sigma^{\Upsilon(5S)}_{b\bar{b}}$ & $ 4.8$ & $ 4.8$ & $ 4.8$ & $ 4.8$\\
$L_{\rm int}$ & $  1.3$ & $  1.3$ & $  1.3$ & $  1.3$\\
$f_{s}$ & $  13.3$ & $  13.3$ & $  13.3$ & $  13.3$ \\
$f_{B_{s}^{*}\bar{B}_{s}^{*}}$ & $ 4.8$ & $ 4.8$ & $ 4.8$ & $ 4.8$\\
Signal MC statistics & $ 0.4$ & $ 0.5$ & $ 0.5$ & $ 0.6$ \\
\hline
Total & $ 19.5$ & $ 24.3$ & $ 25.0$ & $ 20.7$ \\
\hline \hline
\end{tabular}
\end{table}

The second type of systematic uncertainty is determined as follows. 
For the $\mathcal R$
requirement, we use the decay $B^0_s\to D_s^-\pi^+$ to estimate the discrepancy
between data and MC. The same event selection except the continuum 
suppression used in Ref.~\cite{dspi} is applied to reconstruct 
$B^0_s\to D_s^-\pi^+$ candidates, where the $D^-_s$ meson is identified via 
the $D^-_s\to \phi \pi^-, D^-_s\to K^0_s K^-$ and $D^-_s\to K^{*0}_s K^-$ 
decays. When forming the variable $\cal {R}$, the $D^-_s$ mesons are treated as
stable particles to mimic the $B^0_s\to hh$ events and the same sets of weighting 
factors used to combine the modified Fox-Wolfram moments in the $hh$ analysis 
are adopted. We compare the reduction fractions in the  $D^-_s\pi^+$ data and 
MC with the $\cal R$ requirements for the four $hh$ modes to obtain the 
systematic uncertainty. The data-MC differences with various $\mathcal R$ 
requirements are all less than $2/3\sigma$ and we 
conservatively assign the quadratic sum of the data-MC difference and the 
statistical uncertainty on the $ D_s^-\pi^+$ sample as the systematic 
uncertainty. 

The identification of kaons and pions is calibrated using a control sample of  
$D^{*+}\to D^0(K^- \pi^+) \pi^+$ decays. For two-body  $B_s^0\to hh$ decays, 
 this systematic uncertainty is 0.7\% per kaon and 0.6\% per pion. 
The $K_{S}^{0}$ reconstruction efficiency is verified using a sample of 
$D^{+} \to K_{S}^{0}\pi^{+}$ and $D^{+}\to K^{-}\pi^{+}\pi^{+}$ decays. We 
compare the ratio of the yields of the two decay modes with the Monte Carlo 
expectation, which is obtained by generating a large Monte Carlo sample with
the proper continuum and $B\bar{B}$ fractions. A systematic error of 4.9\% per 
$K^0_S$ meson is obtained by adding, in quadrature, the deviation of the data 
and MC ratios
 and the uncertainties of the branching fractions of the two decay modes, where the latter is 
  the dominant error.  
The systematic uncertainty due to the track reconstruction efficiency is estimated using 
partially reconstructed $D^{*}$ events~\cite{Dstar} and is $1\%$ per track. 
Sources of uncertainty in the number of $B_s^*\bar{B}_s^*$ pairs include  $L_{\rm int}$, $\sigma^{\Upsilon(5S)}_{b\bar{b}}$, 
$f_s$, and $f_{B_{s}^{*}\bar{B}_{s}^{*}}$.
Systematic uncertainties are summarized in Table~\ref{sys1}.

\begin{figure}[t!]
\begin{center}
\hspace{-0.2cm}
\includegraphics[width= 4.2cm]{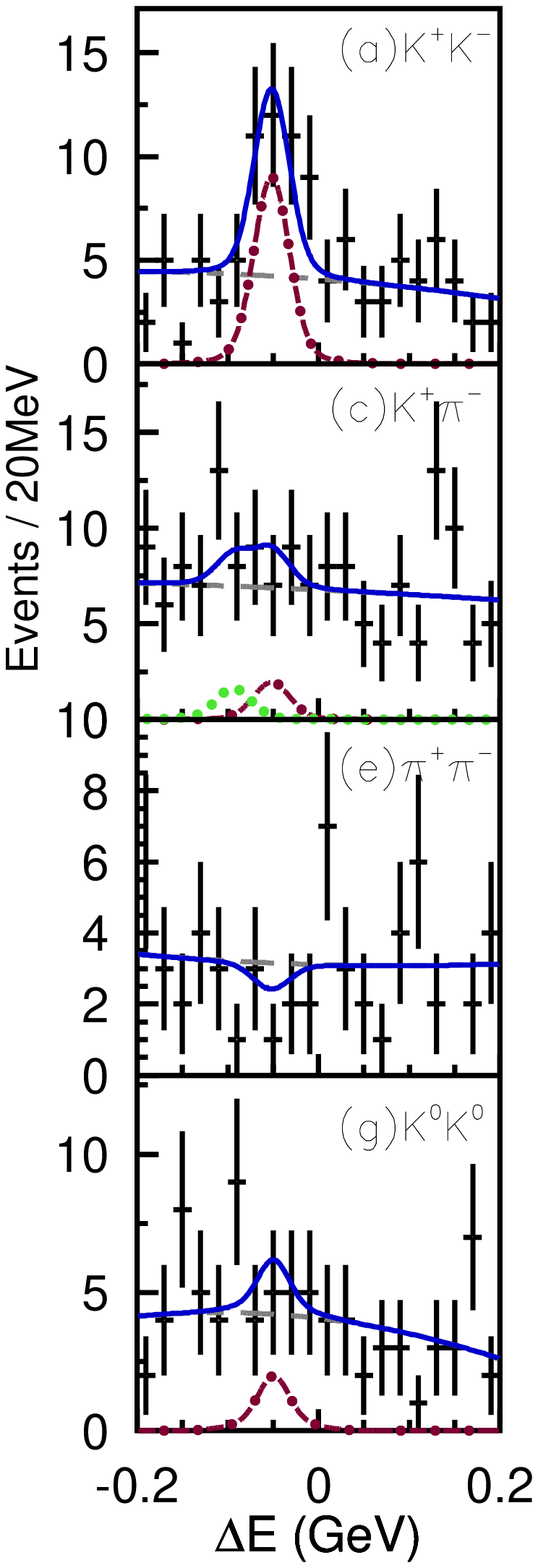}
\hspace{-0.5cm}
\includegraphics[width= 4.2cm]{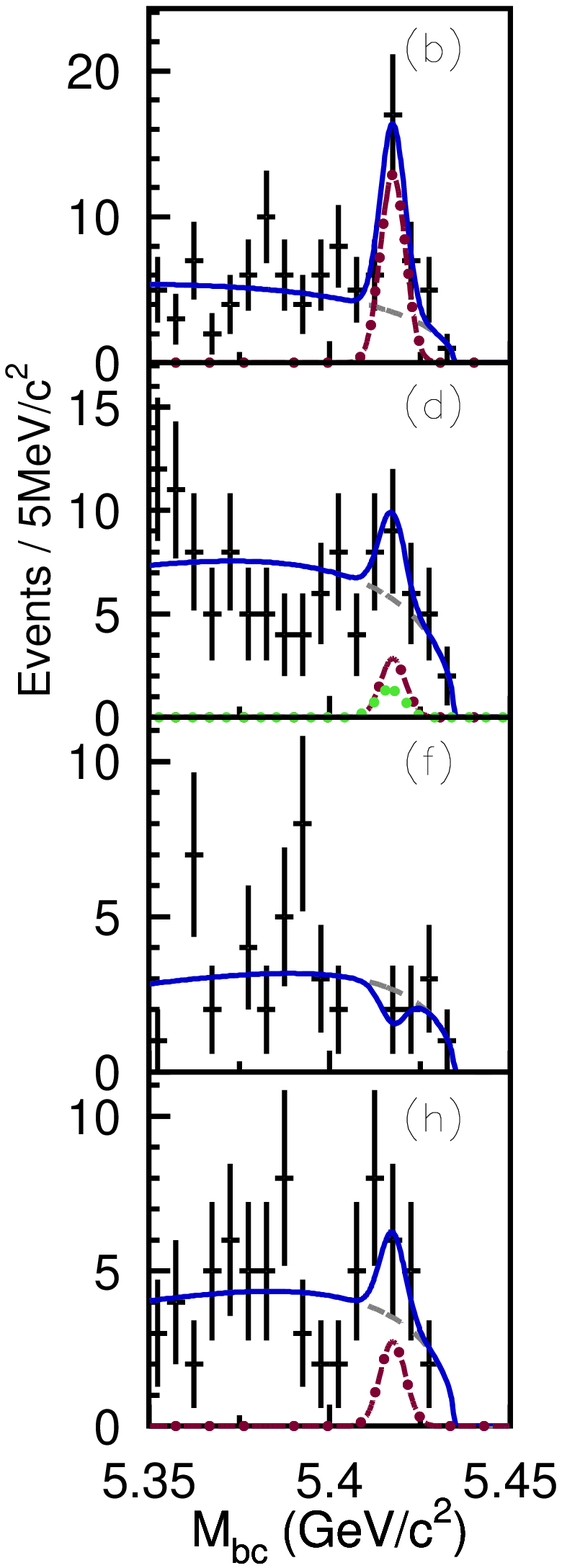}\\
\end{center}
\caption{
Distributions of $\Delta E$ ($M_{\rm bc}$) with fit results superimposed for
the $K^+K^-$ (a,b), $K^+\pi^-$ (c,d), $\pi^+\pi^-$ (e,f), and $K^0\bar{K}^0$ (g,h) events 
in the $M_{\rm bc}$ ($\Delta E$) signal region.
The blue solid curves represent the fit results, in which the red dot-dashed (grey dashed) curves represent signal (continuum background).
The green dotted curves in the $K^{-}\pi^{+}$ plot represent the $K^{+}K^{-}$ cross-feed. 
}
\label{fig:result}
\end{figure}

The fit results are shown in Figure~\ref{fig:result} and summarized in Table~\ref{tab:results}.
A significant signal is observed
in the $B^0_s\to K^+K^-$ mode, and the branching fraction is measured to be
$\mathcal{B} 
   = [3.8_{-0.9}^{+1.0}(\mathrm{stat})\pm 0.5(\mathrm{syst})\pm0.5(f_s)] \times 10^{-5}$ with a significance of $5.8$$\sigma$.
The signal significance is defined by 
$\Sigma = \sqrt{2\ln(\mathcal{L}_{\mathrm{max}}/\mathcal{L}_{0})}$,
where $\mathcal{L}_{\mathrm{max}} (\mathcal{L}_{0})$ is the likelihood 
value at its maximum (with zero signal yield) obtained after convolving the
likelihood function  with
a Gaussian function having  width equal to the fitting systematic uncertainty. 
For the other decay modes, 
the 90\% upper limit ($\mathcal{B}_{90\%}$) is computed as
\begin{equation}
\frac{\int_0^{\mathcal{B}_{90\%}}\mathcal{L}(\mathcal{B})d\mathcal{B}}{\int_0^1\mathcal{L}(\mathcal{B})d\mathcal{B}} = 0.9 ,
\end{equation}
with the likelihood function after convolving with a Gaussian
width equal to the total systematic uncertainty.

\begin{table}[htb]
\caption{Summary of the signal yields, significances ($\Sigma$), reconstruction efficiencies ($\epsilon$), 
branching fractions $(\mathcal{B})$ and upper limits (U.L.) at the $90\%$ confidence level.}
\label{tab:results}
\begin{tabular}
{@{\hspace{0.1cm}}l@{\hspace{0.2cm}}  @{\hspace{0.2cm}}ccccc@{\hspace{0.1cm}}}
\hline \hline
Mode & Yield & $\Sigma$ & $\epsilon$(\%) & $\mathcal{B}$($10^{-5}$) & U.L.($10^{-5}$) \\
\hline
$K^{+}K^{-}$ & $23.4_{-6.3}^{+5.5}$ & 5.8 & 24.5 & $3.8^{+1.0}_{-0.9} \pm 0.5 \pm 0.5$
& $-$\\
$K^{-}\pi^{+}$ & $5.4^{+5.1}_{-4.3}$ &	 1.2  & 21.0    & $-$ &  2.6\\
$\pi^{+}\pi^{-}$ & $-2.0^{+2.3}_{-1.5}$  & 	$-$	&	14.4	& $-$ &	 1.2\\
$K^{0}\bar{K}^{0}$ & $5.2^{+5.0}_{-4.3}$  &1.2 &   8.0   & $ -$ & 6.6\\     

\hline \hline
\end{tabular}
\end{table}

   In conclusion, we observe  $B_{s}^{0} \to K^{+}K^{-}$
 with  
\begin{eqnarray}
&\mathcal{B}(B_s^0\to K^+K^-)\nonumber\\
\ \  =& [3.8_{-0.9}^{+1.0}(\mathrm{stat})\pm 0.5(\mathrm{syst})\pm0.5(f_s)] \times 10^{-5}.  
\end{eqnarray}
Our result is consistent 
with the Standard Model prediction~\cite{PRL97061801} and the CDF measurement ($[2.44\pm 0.14\pm 0.46] \times 10^{-5}$)~\cite{cdf}. 
No significant signals are observed in the other modes, and we set upper limits at 90\% 
confidence level: 
\begin{eqnarray}
\nonumber
\mathcal{B}(B_{s}^{0} \to K^{-}\pi^{+}) < 2.6 \times 10^{-5},\\
\nonumber
\mathcal{B}(B_{s}^{0} \to \pi^{+}\pi^{-}) < 1.2 \times 10^{-5} ,\\
\mathcal{B}(B_{s}^{0} \to K^{0}\bar{K}^{0}) < 6.6 \times 10^{-5}.
\end{eqnarray}
The first two limits are consistent with results from CDF~\cite{cdf2}, although with less sensitivity,
and the third is a first report: this decay is very challenging to reconstruct 
at a hadron collider.







We thank the KEKB group for the excellent operation of the
accelerator, the KEK cryogenics group for the efficient
operation of the solenoid, and the KEK computer group and
the National Institute of Informatics for valuable computing
and SINET3 network support.  We acknowledge support from
the Ministry of Education, Culture, Sports, Science, and
Technology (MEXT) of Japan, the Japan Society for the 
Promotion of Science (JSPS), and the Tau-Lepton Physics 
Research Center of Nagoya University; 
the Australian Research Council and the Australian 
Department of Industry, Innovation, Science and Research;
the National Natural Science Foundation of China under
contract No.~10575109, 10775142, 10875115 and 10825524; 
the Department of Science and Technology of India; 
the BK21 and WCU program of the Ministry Education Science and
Technology, the CHEP SRC program and Basic Research program (grant No.
R01-2008-000-10477-0) of the Korea Science and Engineering Foundation,
Korea Research Foundation (KRF-2008-313-C00177),
and the Korea Institute of Science and Technology Information;
the Polish Ministry of Science and Higher Education;
the Ministry of Education and Science of the Russian
Federation and the Russian Federal Agency for Atomic Energy;
the Slovenian Research Agency;  the Swiss
National Science Foundation; the National Science Council
and the Ministry of Education of Taiwan; and the U.S.\
Department of Energy.
This work is supported by a Grant-in-Aid from MEXT for 
Science Research in a Priority Area (``New Development of 
Flavor Physics"), and from JSPS for Creative Scientific 
Research (``Evolution of Tau-lepton Physics").




%


\begin{thebibliography}{99}
\bibitem{qcon}{
The inclusion of charge-conjugate modes is implied throughout this paper unless explicitly stated.
}
\bibitem{eeqq} The  cross section ($\sigma$) of light quark pair production, 
$e^+e^-\to q \bar{q}$, is estimated using the leading-order 
calculation, $\sigma = \frac{N_c Q_f^2 4\pi \alpha^2}{3s} \beta [1+ \frac{1-\beta^2}{2}]$, where $N_c$ is the number of colors, $Q_f$ is the charge of the 
quark, 
$\alpha$ is the fine structure constant, $s$ is the total enegy squared, and 
$\beta$ is velocity of the quark in the center of mass frame 
divided by the speed of light. The value of 2.446 nb is the cross section sum 
for the four light quark pairs.
\end{thebibliography}
\end{document}